\documentclass[preprint,%
secnumarabic,%
tightenlines,%
superscriptaddress,%
floats,prd,eqsecnum,nobibnotes%
,nofootinbib,aps,letterpaper,11pt]{revtex4}

\usepackage{palatino}
\usepackage{amsmath, amssymb}
\usepackage{xspace}
\usepackage{graphicx}
\usepackage[breaklinks=true,%
            plainpages=false,%
            bookmarks=true
           ]{hyperref}

\usepackage{color} 

\usepackage{subfigure}

\allowdisplaybreaks[3]
\newcommand\oldappendix{This will give an error if \backspace oldappendix
   already exists.}
\let\oldappendix\appendix
\renewcommand\appendix{\oldappendix%
   \renewcommand\theequation{\thesection.\arabic{equation}}}

\newcommand\bigO{O} 

\newcommand\be{\begin{equation}}
\newcommand\bea{\begin{eqnarray}}
\newcommand\ee{\end{equation}}
\newcommand\eea{\end{eqnarray}}
\newcommand\h{\frac{1}{2}}

\newcommand{\bdm}{\begin{displaymath}}
\newcommand{\edm}{\end{displaymath}}

\newcommand\p{\partial}

\newcommand\drm{\mathrm{d}}

\newcommand{\pd}{\partial}

\newcommand{\ints}{\mathbb{Z}}
\newcommand{\veps}{\varepsilon} 

\newcommand{\sech}{\text{\sech}}

\newcommand{\com}[2]{\left[#1,\, #2\right]}
\newcommand{\ac}[2]{\{#1,\,#2\}}

\newcommand{\dg}{\dagger}

\newcommand\ket[1]{| #1 \rangle}

\newcommand\normalTag{\addtocounter{equation}{1}\tag*{\normalsize(\theequation)}}

\makeatletter 
\newenvironment{calc}{
	\start@align\@ne\st@rredtrue\m@ne}
		     {\normalTag\endalign}
\makeatother 

\begin{document}

\title{\vspace*{\fill}Intertwining Relations for the Deformed D1D5 CFT}
\author{Steven G. Avery}\email{avery@mps.ohio-state.edu}
\affiliation{%
      Department of Physics \\
      The Ohio State University \\
      191 West Woodruff Avenue \\
      Columbus, Ohio \ 43210-1117\\ 
      USA
      \vspace*{\fill}}
\author{Borun D. Chowdhury\vspace*{\fill}}\email{borundev@mps.ohio-state.edu}
\affiliation{%
Instituut voor Theoretische Fysica \\
Universiteit van Amsterdam \\
Valckenierstraat 65 \\
Amsterdam 1018XE, The Netherlands
	\vspace*{\fill}}

\begin{abstract}
\vspace*{\baselineskip}

The Higgs branch of the D1D5 system flows in the infrared to a
two-dimensional $\mathcal{N}=(4,4)$ SCFT.  This system is believed to
have an ``orbifold point'' in its moduli space where the SCFT is a
free sigma model with target space the symmetric product of copies of
$T^4$; however, at the orbifold point gravity is strongly coupled and
to reach the supergravity point one needs to turn on the four exactly
marginal deformations corresponding to the blow-up modes of the
orbifold SCFT.  Recently, technology has been developed for studying
these deformations and perturbing the D1D5 CFT off its orbifold point.
We present a new method for computing the general effect of a single
application of the deformation operators.  The method takes the form
of intertwining relations that map operators in the untwisted sector
before application of the deformation operator to operators in the
2-twisted sector after the application of the deformation operator.
This method is computationally more direct, and may be of theoretical
interest. This line of inquiry should ultimately have relevance for
black hole physics.

\vspace*{\fill}
\end{abstract}

\maketitle

\section{Introduction}

The D1D5 system has been a very rich area for exploring black hole
physics: one can account for the black hole entropy by counting states
of the dual CFT~\cite{sen1, sen2, stromvafa}; one can compute the
spectrum and rate of absorption and emission from the gravitational
description and match it to a CFT description~\cite{radiation-1,
  radiation-2, radiation-3, radiation-4, radiation-5, callan96,
  emparan,cv,cm1,cm2,cm3,acm1,ac1}; and one can create and study
explicit microstate geometries (see~\cite{fuzzballs2-1, fuzzballs2-2,
  fuzzballs2-3, fuzzballs2-4, BalaDeBoer} for reviews of older work,
and~\cite{bena10, bena-bolt, bena-branes, deBoer10} for some recent
developments).

There is believed to be an orbifold point in the D1D5 moduli space
where a free two-dimensional CFT with orbifolded target space becomes
a good description~\cite{swD1D5, fmD1D5, deBoerD1D5, dijkgraafD1D5,
  frolov-1, frolov-2, Jevicki}.  Because calculations are easier at
the orbifold point, all of the above mentioned CFT calculations are
done there. At the orbifold point, however, the gravity approximation
is not valid. While the entropy-counting calculations are protected on
account of supersymmetry, the agreement of so many absorption and
emission calculations at the orbifold point with supergravity
calculations at a different point in moduli space strongly suggests
some kind of ``non-renormalization'' theorem~\cite{dmw02}.

To understand more complicated black hole physics, it becomes
necessary to deform the CFT off of the orbifold point by blowing up
the fixed points of the orbifold. We do this by introducing four
marginal deformation operators.\footnote{The four deformation
  operators form a 2-index tensor of $SU(2)_2$, so we sometimes call
  this object ``the deformation operator.''} While the supergravity
point in moduli space is far from the orbifold point, we take a
perturbative approach and start by looking at the action of a single
application of the deformation operator. We hope that this approach
will yield insight into black hole physics. To this end, \cite{acm2}
computes the effect of the deformation operator on the vacuum,
and~\cite{acm3} shows an algorithm for computing the effect of the
deformation operator on more general states of the CFT.

In this paper, we introduce a new and more direct method for computing
the effect of the deformation operator on general states. This method
may also be of theoretical interest, because it makes the physics of
twist operators closer to that of Bogolyubov transformations. We start
by looking for a Bogolyubov-type relation between operators ``before''
the deformation and ``after'' the deformation. Unfortunately, the
method suffers from multidimensional series whose value depends
sensitively on the way one evaluates them. We resolve these
ambiguities by introducing a physically motivated prescription for
evaluating the infinite series; however, the prescription makes our
relation a weaker statement, since the prescription depends on what
other excitations one has. Thus, we term the connection between
operators before the deformation and after the deformation
``intertwining relations'' instead of Bogolyubov coefficients.

In Section~\ref{sec:CFT}, we briefly introduce the deformation
operator for the D1D5 CFT and review results of~\cite{acm2}. In
particular, the deformation operator contains a 2-twist operator,
whose effect on the vacuum is to produce a ``squeezed state.'' The
squeezed state, written as an exponential of pairs of bosonic and
fermionic modes acting on the twisted vacuum, is reminiscent of the
state one generically gets from a Bogolyubov transformation of the
vacuum. The twist operator is the only difficult part of working with
the deformation operator, and for this paper we restrict our attention
to it.

In Section~\ref{sec:intertwining}, we make the above analogy more
precise by introducing the analogue of the Bogolyubov coefficients,
that is we show how the 2-twist operator maps modes of the untwisted
sector into twisted sector modes. We first give a simple, formal
derivation of these intertwining coefficients. Then we give an example
of the order-of-summation issue that results from using the
coefficients. We can understand the problem by making a more rigorous
derivation of the intertwining relations, which allows us to give a
prescription that gives the correct answer. In~\ref{sec:prescription}
we concisely state the prescription.

In Section~\ref{sec:example}, we demonstrate the prescription and the
validity of the more rigorous derivation by computing a particular
state in two distinct ways and showing agreement.

Finally, we recapitulate the main points and discuss future directions
in the conclusion. Our notation and conventions for the CFT are
outlined in Appendix~\ref{ap:CFT-notation}. Appendix~\ref{ap:series}
contains a list of series that we use. Some of them we could not prove
directly, but we are confident are correct after numerical
comparisons.

\section{The Deformation Operator}\label{sec:CFT}

Recall that the D1D5 system is IIB string theory compactified
on\footnote{One could consider also K3 instead of $T^4$; however, for
  simplicity we consider only $T^4$.}
\begin{equation}
M_{9,1}\rightarrow M_{4,1}\times S^1\times T^4,
\end{equation}
with a bound state of $N_1$ D1 branes wrapping $S^1$ and $N_5$ D5
branes wrapping $T^4\times S^1$. For $S^1$ large compared to the
$T^4$, the low energy description of the resulting bound state is a
1+1-dimensional CFT. The CFT has $SU(2)_L\times SU(2)_R$ R-symmetry
that corresponds to the isometry of the near-horizon $S^3$.
Additionally, we find it convenient to label fields with the $SO(4)_I
\simeq SU(2)_1\times SU(2)_2$ symmetry corresponding to the torus
directions~\cite{Jevicki}. This symmetry is broken by the torus, but is still useful
to organize the fields.

It has been conjectured that we can move to a point in the moduli
space of the D1D5 system called the ``orbifold point,'' where the CFT
is a 1+1-dimensional sigma model with target space
$(T^4)^{N_1N_5}/S_{N_1N_5}$~\cite{swD1D5, fmD1D5,deBoerD1D5,
  dijkgraafD1D5,frolov-1,frolov-2, Jevicki}. Each of the $N_1N_5$
copies consist of 4 real bosonic fields $X$ and four real fermionic
fields $\psi$, see Appendix~\ref{ap:CFT-notation} and~\cite{acm2,
  acm3} for more details.  We parameterize the base space with one
complex coordinate $z$, using radial quantization.

The orbifold theory has different twist sectors, which correspond to
circling $S^1$ and coming back only up an element of the orbifold
group, $S_{N_1N_5}$. The twist operators $\sigma_n$ map between
different twist sectors. We do not need most of the technical details
of the twist operators in this paper, but we do use the fact that one
can map to a covering space which looks like the untwisted sector of
the theory via a meromorphic mapping $z=z(t)$~\cite{lm1}.

For this paper, we only need the chiral primary 2-twist operator
$\sigma_2^{++}(z)$, where the plusses indicate that it is the top
member of both a $SU(2)_L$ doublet and a $SU(2)_R$ doublet. If the
only twist operators are a $\sigma_2^{++}(z_0)$ and another 2-twist at
$z=\infty$, then
\begin{equation}
z = z_0 + t^2
\end{equation}
maps to a single-valued covering space parameterized by the complex
coordinate $t$. This map removes the twisting, but the $SU(2)_L\times
SU(2)_R$ charge is preserved by inserting spin fields,
{$S^+(t=0)\bar{S}^+(\bar{t}=0)$}, which also ensure the correct
fermion boundary conditions~\cite{lm2}.

The D1D5 CFT has 20 exactly marginal deformation
operators~\cite{dmw99, dmw02,gomis,gava} that correspond to different
directions in moduli space.  Four of those operators are the blow-up
modes of the orbifold. These are the deformation operators that move
us to the supergravity point in the moduli space and hence the ones
which we are most interested in.

\subsection{The operator}

The deformation operator is a singlet under $SU(2)_L\times SU(2)_R$.
To obtain such a singlet we apply modes of $G^\mp_{\dot A}$ to
$\sigma_2^\pm$. In~\cite{acm2} it was shown that we can write the
deformation operator(s) as
\begin{equation}\label{eq:def-op}
\widehat O_{\dot A\dot B}(w_0)=
  \Big[\int_{w_0} \frac{\drm w}{2\pi i} G^-_{\dot A} (w)\Big]
  \Big[\int_{\bar w_0} \frac{\drm \bar{w}}{2\pi i} \bar G^-_{\dot B} (\bar w)\Big]
   \sigma_2^{++}(w_0)
\end{equation}
The left and right movers separate out completely for all the
computations that we perform. Thus from now on we work with the left
movers only; in particular, we write the twist operator only with its
left spin: $\sigma_2^+$.

The operator $\sigma^+_2$ is normalized to have a unit OPE with its
conjugate
\begin{equation}
\sigma_{2,+}(z')\sigma_2^{+}(z)\sim {1\over (z'-z)}.
\end{equation}
This implies that acting on the Ramond vacuum~\cite{acm2}
\begin{equation}
\sigma_2^+(z)|0_R^-\rangle^{(1)} |0_R^-\rangle^{(2)}=|0_R^-\rangle+\bigO(z).
\label{qwthree}
\end{equation}
Here $|0_R^-\rangle$ is the spin down Ramond vacuum of the CFT on the
doubly wound circle produced after the twist. The normalization
\eqref{qwthree} has given us the coefficient unity for the first term
on the RHS, and the $\bigO(z)$ represent excited states of the
CFT on the doubly wound circle.

In this paper, we focus on the effect of the twist operator since the
effect of the susy-current, $G$, is easy to compute: one can break the
$G$-contours in Equation~\eqref{eq:def-op} into contours acting
before the twist operator and contours acting after the twist
operator~\cite{acm2, acm3}.

\subsection{The action of the twist operator on the vacuum}

Since we are mainly interested in better understanding the twist
operator, it is useful to define
\begin{equation}
\ket{\chi} = \sigma_2^+(z_0)\ket{0_R^-}^{(1)}\ket{0_R^{-}}^{(2)},
\end{equation}
and recall that in~\cite{acm2} it was found that the twist operator
acting on the Ramond vacuum may be written as the squeezed state
\begin{equation}
\ket{\chi} = \exp\left[
  -\frac{1}{2}\sum_{m, n}\gamma^B_{mn}\epsilon^{AB}\epsilon^{\dot{A}\dot{B}}
                                             \alpha_{A\dot{A},-m}\alpha_{B\dot{B},-n}
+\sum_{m,n}\gamma^F_{mn}\epsilon_{AB}\psi^{+A}_{-m}\psi^{-B}_{-n}\right]\ket{0^-_R}
\label{pfive}
\end{equation}
where
\begin{subequations}
\begin{align}\label{gammaB}
\gamma^B_{mn} &= \begin{cases}
\frac{4 z_0^{\frac{m+n}{2}}}{mn(m+n)\pi}\frac{\Gamma(\frac{m}{2}+1)\Gamma(\frac{n}{2}+1)}{\Gamma(\frac{m+1}{2})\Gamma(\frac{n+1}{2})} & m,n\,\text{odd, positive}\\
0 & \text{otherwise}\end{cases}
\\
\gamma^F_{mn} &= \begin{cases}
-\frac{2z_0^\frac{m+n}{2}}{n(m+n)\pi}\frac{\Gamma(\frac{m}{2}+1)\Gamma(\frac{n}{2}+1)}{\Gamma(\frac{m+1}{2})\Gamma(\frac{n+1}{2})}
 & m, n\,\text{odd, positive}\\
0 & \text{otherwise}.\end{cases}
\end{align}
\end{subequations}

\section{Intertwining Relations}\label{sec:intertwining}

We first present a formal, intuitive derivation of the intertwining
relations, which relate modes of the untwisted sector to modes of the
twisted sector. In Section~\ref{sec:subtleties}, we show how the
derived relations lead to multi-dimensional infinite series whose
value depends on how one takes the limit of the partial sums going to
infinity. We then give a more rigorous derivation of the relations,
which suggests a prescription on how to evaluate the ambiguous series.

\subsection{Basic Derivation}

We are interested in finding a relationship between modes in the
untwisted sector, ``before the twist,'' and modes in the twisted
sector, ``after the twist.'' More precisely we would like to know in
general how to write
\begin{calc}
\sigma^+_2(z_0) \big(\text{excitations before the twist}\big)\ket{0_R^-}^{(1)}\ket{0_R^-}^{(2)} 
  &= \big(\text{excitations after the twist}\big)\sigma^+_2(z_0)\ket{0_R^-}^{(1)}\ket{0_R^-}^{(2)}\\
  &= \big(\text{excitations after the twist}\big)\ket{\chi}.
\end{calc}
In~\cite{acm3}, an algorithm was found for doing just that; however,
the question this paper addresses is whether there is a more general
relation between individual modes before the twist operator and after
the twist operator.

Since our argument does not depend on $SU(2)_L$ or
$SU(2)_1\times SU(2)_2$, we suppress the indices on the bosons and
fermions. We begin by noting that before the twist the correct field
expansions are
\begin{equation}\label{eq:bosons-before}
i\pd X^{(1)}(z) = \sum_n \frac{\alpha^{(1)}_n}{z^{n+1}}\qquad
i\pd X^{(2)}(z) = \sum_n \frac{\alpha^{(2)}_n}{z^{n+1}}\qquad
|z|<|z_0|,
\end{equation}
and
\begin{equation}\label{eq:fermions-before}
\psi^{(1)}(z) = \sum_{n\in\ints + \frac{1}{2}} \frac{\psi^{(1)}_n}{z^{n+\frac{1}{2}}}\qquad
\psi^{(2)}(z) = \sum_{n\in\ints + \frac{1}{2}} \frac{\psi^{(2)}_n}{z^{n+\frac{1}{2}}}\qquad
|z|<|z_0|.
\end{equation}
After the twist, the correct expansions are given by
\begin{equation}\label{eq:bosons-after}
i\pd X^{(1)}(z) = \frac{1}{2}\sum_n \frac{\alpha_n}{z^{\frac{n}{2}+1}}\qquad
i\pd X^{(2)}(z) = \frac{1}{2}\sum_n \frac{(-1)^n\alpha_n}{z^{\frac{n}{2}+1}}\qquad
|z|>|z_0|,
\end{equation}
and
\begin{equation}\label{eq:fermions-after}
\psi^{(1)}(z) = \frac{1}{2}\sum_n\frac{\psi_n}{z^{\frac{n}{2}+\frac{1}{2}}}\qquad
\psi^{(2)}(z) = \frac{1}{2}\sum_n\frac{(-1)^n\psi_n}{z^{\frac{n}{2}+\frac{1}{2}}}\qquad
|z|>|z_0|.
\end{equation}
Note that after the twist, there is no unique way of distinguishing
copies $(1)$ and $(2)$, but the above expansions correspond to
\emph{a} way of defining $(1)$ and $(2)$. Following~\cite{acm2,acm3},
we define the modes in the twisted sector with an extra factor of 2 so
that we can work with integers.

Now, recall that the fields $\pd X(z)$ and $\psi(z)$ should be
holomorphic functions except at isolated points where there are other
operator insertions. In particular, there is nothing special that
occurs on the circle $|z|=|z_0|$ (there \emph{is} something special at
the isolated point $z_0$). The curve is the boundary between our
twisted and untwisted mode expansions about the origin $z=0$, but if
one were to do mode expansions about a different point in the complex
plane then the expansions would change across a different curve (that
still passes through $z_0$).  Therefore, \emph{at least away from the
  twist operator at $z_0$}, we expect that the fields $\pd X(z)$ and
$\psi(z)$ should be continuous across $|z|=|z_0|$.

Thus, on the circle $|z|=|z_0|$ ({\em  excluding some neighborhood around
$z=z_0$}) we may identify
\begin{equation}\label{eq:boson-exp-equality}
\sum_n \frac{\alpha^{(1)}_n}{z^{n+1}} = \frac{1}{2}\sum_n \frac{\alpha_n}{z^{\frac{n}{2}+1}}
\qquad
\sum_n \frac{\alpha^{(2)}_n}{z^{n+1}} = \frac{1}{2}\sum_n \frac{(-1)^n\alpha_n}{z^{\frac{n}{2}+1}},
\qquad
|z|=|z_0|
\end{equation}
and
\begin{equation}\label{eq:fermion-exp-equality}
\sum_{n\in\ints + \frac{1}{2}} \frac{\psi^{(1)}_n}{z^{n+\frac{1}{2}}}
 = \frac{1}{2}\sum_n\frac{\psi_n}{z^{\frac{n}{2}+\frac{1}{2}}}
\qquad
\sum_{n\in\ints + \frac{1}{2}} \frac{\psi^{(2)}_n}{z^{n+\frac{1}{2}}}
  = \frac{1}{2}\sum_n\frac{(-1)^n\psi_n}{z^{\frac{n}{2}+\frac{1}{2}}},
\qquad
|z|=|z_0|.
\end{equation}
Multiplying \eqref{eq:boson-exp-equality} by $z^m$ and integrating
along the circle $|z|=|z_0|$, we
get
\begin{calc}\label{eq:boson-coef}
\alpha_m^{(1)} &= \frac{1}{2}\sum_n\alpha_n \int\frac{dz}{2\pi i} z^{m-\frac{n}{2}-1}\\
  &= \frac{1}{2}\sum_n\alpha_n \int_0^{2\pi}\frac{d\theta}{2\pi} (z_0e^{i\theta})^{m-\frac{n}{2}}\\
  &= \frac{1}{2}\alpha_{2m} + \frac{i}{2\pi}
     \sum_{n\,\text{odd}}\frac{z_0^{m-\frac{n}{2}}}{m-\frac{n}{2}}\alpha_n,
\end{calc}
and similarly
\begin{equation}\label{eq:boson-coef-2}
\alpha_m^{(2)} = \frac{1}{2}\alpha_{2m} 
             - \frac{i}{2\pi}\sum_{n\,\text{odd}}\frac{z_0^{m-\frac{n}{2}}}{m-\frac{n}{2}}\alpha_n,
\end{equation}
where the sum over the odds is both positive and negative.  These are
the desired relations between modes before the twist and modes after
the twist. The relations are analogous to the more general Bogolyubov
transformations discussed in condensed matter in~\cite{deGennes}. Note
that the contour in~\ref{eq:boson-coef} is open, since we must exclude
some infinitesimal neighborhood around $z_0$. It is straightforward to
find the analogous relation for fermions in the R sector:
\begin{equation}\label{eq:fermion-coef}
\psi^{(1,2)}_n = 
\frac{1}{2}\psi_{2n} 
  \pm \frac{i}{2\pi}\sum_{k\,\text{odd}}\frac{z_0^{n-\frac{k}{2}}}{n-\frac{k}{2}}\psi_k.
\end{equation}
Given the delicate nature of the above argument, in particular with
regard to what is happening around $z_0$, one may not be surprised
that there are some hidden subtleties with these relations. Note also
that the derivation would seem to work for any holomorphic field
$\mathcal{O}(z)$, which \emph{cannot be correct} since some modes have
a nontrivial commutator with the twist operator, e.g. $J_0^-$. We should be
careful in what we mean when we write ``$=$'' in the above
expressions.  For example in Equation~\ref{eq:boson-coef}, we
implicitly mean the operator relation
\begin{equation}
\sigma_2^+(z_0)\alpha_m^{(1)} = \left[\frac{1}{2}\alpha_{2m} + \frac{i}{2\pi}
     \sum_{n\,\text{odd}}\frac{z_0^{m-\frac{n}{2}}}{m-\frac{n}{2}}\alpha_n\right]\sigma_2^+(z_0)
\end{equation}
with the above radial ordering. The usage should be clear from the
context.

Before showing what goes wrong, let us first explore what goes right.
First of all this is the kind of relation we were hoping for: it
relates positive and negative modes before the twist to positive and
negative modes after the twist directly. The method given
in~\cite{acm3} can only relate states to states; it cannot relate an
individual mode before the twist to modes after the twist without
knowing what other excitations one has before the twist. 

An important requirement for Bogolyubov coefficients is that they
respect the commutation relations. We can check that the above
relations are consistent with the commutation relations by computing,
for instance,
\begin{calc}\label{eq:comm-check}
\com{\alpha_m^{(1)}}{\alpha_n^{(1)}} &=
\com{\frac{1}{2}\alpha_{2m} 
   + \frac{i}{2\pi}\sum_{k\,\text{odd}}\frac{z_0^{m-\frac{k}{2}}}{m-\frac{k}{2}}\alpha_k}
    {\frac{1}{2}\alpha_{2n}
   + \frac{i}{2\pi}\sum_{l\,\text{odd}}\frac{z_0^{n-\frac{l}{2}}}{n-\frac{l}{2}}\alpha_l}\\
 &= \frac{m}{2}\delta_{m+n,0} 
   - \frac{z_0^{m+n}}{4\pi^2}\sum_{k\,\text{odd}}\frac{k}{(m-\frac{k}{2})(n+\frac{k}{2})}.
\end{calc}
The sum is divergent; however, if we make the relatively modest
assumption that we should cutoff the sum symmetrically for positive
and negative $k$, then we find\footnote{This mild UV ambiguity might
  be seen as a hint of the other UV issues with the intertwining
  relations; however, this issue does not arise for the fermions, and
  it is of a different character. The UV issues that we discuss at
  length arise with multidimensional series; whereas the above is
  arguably the only reasonable regularization of the series
  in~\ref{eq:comm-check}.  For example, if one cuts off the positive
  modes at $L$ and the negative modes at $-2L$ then one \emph{does}
  get a different answer, but it is not a consistent truncation of the
  Hilbert space to have a creation operator without its corresponding
  annihilation operator.}
\begin{equation}
\lim_{L\to\infty}\sum_{k\,\text{odd}}^{|k|<L}\frac{k}{(m-\frac{k}{2})(n+\frac{k}{2})}
 = - 2m\pi^2\delta_{m+n,0},
\end{equation}
which gives the correct answer. Similar calculations go through for
the other (anti-)commutations.

Second, it gets the right answer for moving a single mode through the
twist operator. For instance,
\begin{calc}
\sigma^+_2(z_0)\alpha^{(1)}_{A\dot{A},n}\ket{0_R^-}^{(1)}\ket{0_R^-}^{(2)}
 &= \left(\frac{1}{2}\alpha_{A\dot{A},2n} + \frac{i}{2\pi}
    \sum_{k\,\text{odd}}\frac{z_0^{n-\frac{k}{2}}}{n-\frac{k}{2}}\alpha_{A\dot{A},k}\right)
    \sigma^+_2(z_0)\ket{0_R^-}^{(1)}\ket{0_R^-}^{(2)}\\
 &= \left(\frac{1}{2}\alpha_{A\dot{A}, 2n} + \frac{i}{2\pi}
    \sum_{k\,\text{odd}}\frac{z_0^{n-\frac{k}{2}}}{n-\frac{k}{2}}\alpha_{A\dot{A},k}\right)
    \ket{\chi}\\
 &= \left[\frac{1}{2}\alpha_{A\dot{A}, 2n} + \frac{i}{2\pi}
          \left(\sum_{k\,\text{odd}^+} \frac{z_0^{n+\frac{k}{2}}}{n+\frac{k}{2}}\alpha_{A\dot{A},-k}
      + \sum_{k,l\,\text{odd}^+} \frac{z_0^{n-\frac{k}{2}}}{n-\frac{k}{2}}
                              \gamma^B_{kl}k\alpha_{A\dot{A}, -l}\right)\right]\ket{\chi}.\\
\end{calc}
Making use of the identity in~\eqref{eq:id-1} one sees that
\begin{equation}\label{eq:gamma-id}
\sum_{k\,\text{odd}^+} \frac{z_0^{n-\frac{k}{2}}k\gamma^B_{kl}}{n-\frac{k}{2}}
 = \frac{z_0^{n+\frac{l}{2}}}{n+\frac{l}{2}}
      \left(\frac{\Gamma(\frac{l}{2})\Gamma(-n+\frac{1}{2})}{\Gamma(\frac{l+1}{2})\Gamma(-n)} 
              - 1\right),
\end{equation}
and thus
\begin{equation}\label{eq:single-boson}
\sigma^+_2(z_0)\alpha^{(1)}_{A\dot{A},n}\ket{0_R^-}^{(1)}\ket{0_R^-}^{(2)}
 = \left[\frac{1}{2}\alpha_{A\dot{A},2n} + \frac{i}{2\pi}\sum_{l\,\text{odd}^+}
         \frac{z_0^{n+\frac{l}{2}}}{n+\frac{l}{2}}
    \frac{\Gamma(\frac{l}{2})\Gamma(-n+\frac{1}{2})}{\Gamma(\frac{l+1}{2})\Gamma(-n)}
    \alpha_{A\dot{A},-l}
  \right]\ket{\chi}.
\end{equation}
For $n$ positive the above vanishes as it should, since the positive
even mode in the first term annihilates $\ket{\chi}$ and $\Gamma(-n)$
kills the second term. For $n$ negative this reproduces the result
found in~\cite{acm3} for a single mode in the
initial state.  

Similarly, if one performs the analogous calculation for
fermions with~\eqref{eq:fermion-coef}, then one can use
\begin{equation}
\psi^{\alpha A}_{+k}\ket{\chi} = 2\sum_{p\,\text{odd}^+}
                              \left(\gamma^F_{pk}\delta^\alpha_+\psi^{+A}_{-p}
                                -\gamma^F_{kp}\delta^\alpha_-\psi^{-A}_{-p}\right)
                              \ket{\chi}\qquad k\,\text{odd, positive},
\end{equation}
to find
\begin{subequations}\label{eq:single-fermion}
\begin{align}
\sigma_2^+(z_0)\psi^{(1)+A}_{n}\ket{0_R^-}^{(1)}\ket{0_R^-}^{(2)}
 &= \left[\frac{1}{2}\psi^{+A}_{2n} + \frac{i}{2\pi}
        \sum_{p\,\text{odd}^+}\frac{z_0^{n+\frac{p}{2}}}{n+\frac{p}{2}}
           \frac{\Gamma(\frac{p}{2}+1)\Gamma(-n+\frac{1}{2})}{\Gamma(\frac{p+1}{2})\Gamma(-n+1)}
           \psi^{+A}_{-p}\right]\ket{\chi}\\
\sigma_2^+(z_0)\psi^{(1)-A}_{n}\ket{0_R^-}^{(1)}\ket{0_R^-}^{(2)}
 &= \left[\frac{1}{2}\psi^{-A}_{2n} + \frac{i}{2\pi}
        \sum_{p\,\text{odd}^+}\frac{z_0^{n+\frac{p}{2}}}{n+\frac{p}{2}}
           \frac{\Gamma(\frac{p}{2})\Gamma(-n+\frac{1}{2})}{\Gamma(\frac{p+1}{2})\Gamma(-n)}
           \psi^{-A}_{-p}\right]\ket{\chi}.
\end{align}
\end{subequations}
This agrees with~\cite{acm3}.

\subsection{Problems}\label{sec:problems}

There are two problems with the above derivation. One is that this
formal derivation only makes use of the holomorphicity of the fields,
which means that one could make the same argument for any other
holomorphic field. For instance, consider $J^a(z)$, one would get
\begin{equation}
J^{a(1,2)}_n \overset{?}{=} \frac{1}{2}J^a_{2n}
             \pm \frac{i}{2\pi}\sum_{k\,\text{odd}}\frac{z_0^{n-\frac{k}{2}}}{n-\frac{k}{2}} J^a_k,
\end{equation}
which leads to 
\begin{calc}\label{eq:wrong-J-com}
\com{J^-_0}{\sigma_2^+(z_0)} &= J^-_0\sigma_2^+(z_0) 
              - \sigma_2^+(z_0)\big(J_0^{(1)} + J_0^{(2)}\big)\\
 &\overset{?}{=} 0.
\end{calc}
One should find $\sigma_2^-(z_0)$, not zero. One finds similar
contradictions if one tries to use the same argument for $T(z)$, too.

The second problem, alluded to above, concerns using the intertwining
relations with more than one mode. The simplest instance may be to
consider
\begin{equation}
\sigma_2^+(z_0)\alpha^{(1)}_{++,m}\alpha^{(1)}_{--,-n}\ket{0_R^-}^{(1)}\ket{0_R^-}^{(2)}
 = -m\delta_{m,n}\ket{\chi}\qquad m,n>0.
\end{equation}
If we use Equation~\ref{eq:boson-coef}, then we get
\begin{calc}\label{eq:two-boson-problem}
- m \delta_{m,n}\ket{\chi}
 &\overset{?}{=}
\left(\frac{1}{2}\alpha_{++, 2m} 
        + \frac{i}{2\pi}\sum_{k\,\text{odd}}\frac{z_0^{m-\frac{k}{2}}}{m-\frac{k}{2}}
            \alpha_{++, k}\right)
\left(\frac{1}{2}\alpha_{--, -2n} 
        + \frac{i}{2\pi}\sum_{l\,\text{odd}}\frac{z_0^{-n-\frac{l}{2}}}{-n-\frac{l}{2}}
            \alpha_{--,l}\right)\ket{\chi}\\
 &= -\frac{m}{2}\delta_{m,n}\ket{\chi}
    -\frac{1}{4\pi^2}\left(\sum_{k\,\text{odd}}\frac{z_0^{m-\frac{k}{2}}}{m-\frac{k}{2}}
                          \alpha_{++, k}\right)
                   \left(\sum_{l\,\text{odd}}\frac{z_0^{-n-\frac{l}{2}}}{-n-\frac{l}{2}}
                         \alpha_{--,l}\right)\ket{\chi}\\
 &= -\frac{m}{2}\delta_{m,n}\ket{\chi}
    -\frac{1}{4\pi^2}\left(\sum_{k\,\text{odd}}\frac{z_0^{m-\frac{k}{2}}}{m-\frac{k}{2}}
                          \alpha_{++, k}\right)
                   \left[\sum_{j\,\text{odd}^+}\alpha_{--,-j}\left(
                       \frac{z_0^{-n+\frac{j}{2}}}{-n+\frac{j}{2}} + \sum_{l\,\text{odd}^+}
                       \frac{z_0^{-n-\frac{l}{2}}l\gamma_{lj}}{-n-\frac{l}{2}}
                         \right)\right]\ket{\chi}\\
\end{calc}
Note that the even--odd cross-terms vanish since they commute and
either the $\alpha_{++,2m}$ or the $\alpha_{++,k}$-sum kills
$\ket{\chi}$ (from~\eqref{eq:single-boson}). Similarly, we need only
look at the commutator of the $\alpha_{++,k}$-sum and the
square-bracketed expression, which gives
\begin{calc}\label{eq:ambig-sum}
&-\left(\sum_{k\,\text{odd}^+}\frac{kz_0^{m-\frac{k}{2}}}{m-\frac{k}{2}}\right)
 \left(\frac{z_0^{-n+\frac{k}{2}}}{-n+\frac{k}{2}} + \sum_{l\,\text{odd}^+}
                       \frac{z_0^{-n-\frac{l}{2}}l\gamma_{lk}}{-n-\frac{l}{2}}
                         \right)\\
&= z_0^{m-n}\sum_{k\,\text{odd}^+}\left(\frac{k}{(m-\frac{k}{2})(n-\frac{k}{2})}
  + \sum_{l\,\text{odd}^+}\frac{z_0^{-\frac{k}{2}-\frac{l}{2}}kl\gamma_{kl}}
                                          {(m-\frac{k}{2})(n+\frac{l}{2})}\right).
\end{calc}
We would like the above expression to evaluate to $2\pi^2 m
\delta_{m,n}$ in order to get the correct answer; however, the above
summations depend sensitively on the order in which one adds the
infinite number of terms. For instance, if we attempt to perform the
$k$-sum \emph{first}, then the first term is divergent and the second
term, using~\eqref{eq:gamma-id}, gives
\begin{equation}
-\sum_{l\,\text{odd}^+}\frac{l}{(m+\frac{l}{2})(n+\frac{l}{2})},
\end{equation}
which is also divergent. 

On the other hand, if we perform the $l$ sum first,
using~\eqref{eq:gamma-id} we find
\begin{calc}
& z_0^{m-n}\sum_{k\,\text{odd}^+}\left(\frac{k}{(m-\frac{k}{2})(n-\frac{k}{2})}
  + \sum_{l\,\text{odd}^+}\frac{z_0^{-\frac{k}{2}-\frac{l}{2}}kl\gamma_{kl}}
                                          {(m-\frac{k}{2})(n+\frac{l}{2})}\right)\\
&= z_0^{m-n}\sum_{k\,\text{odd}^+}\left(\frac{k}{(m-\frac{k}{2})(n-\frac{k}{2})}
  + \frac{k}{(m-\frac{k}{2})(n-\frac{k}{2})}
     \left[\frac{\Gamma(\frac{k}{2})\Gamma(n+\frac{1}{2})}{\Gamma(\frac{k+1}{2})\Gamma(n)} 
            - 1\right]\right)\\
&= z_0^{m-n}\frac{\Gamma(n+\frac{1}{2})}{\Gamma(n)}
  \sum_{k\,\text{odd}^+}\left(\frac{k}{(m-\frac{k}{2})(n-\frac{k}{2})}
  \frac{\Gamma(\frac{k}{2})}{\Gamma(\frac{k+1}{2})}\right),
\end{calc}
which using another identity,
\begin{equation}
\sum_{k\,\text{odd}^+}\frac{k}{(m-\frac{k}{2})(n-\frac{k}{2})}
      \frac{\Gamma(\frac{k}{2})}{\Gamma(\frac{k+1}{2})}
 = 2\pi^2 m \frac{\Gamma(n)}{\Gamma(n+\frac{1}{2})}\delta_{m,n}\qquad m,n>0,
\end{equation}
gives the correct answer.

How should we think of the ambiguity in Equation~\ref{eq:ambig-sum}?
Any infinite series is implicitly evaluated by determining the limit
of a sequence of \emph{partial sums}. In our case, higher values of
$k$ and $l$ correspond to higher modes, so it is natural to think of
imposing UV cutoffs on the sums, $k<L_1$ and $l<L_2$. We then wish to
take the limit as $L_1, L_2\to\infty$, but there are many different
ways to do that. If we define $b=L_1/L_2$ to parameterize the
different ways of evaluating~\ref{eq:ambig-sum}, then evaluating the
$k$-sum first corresponds to $b=\infty$, while evaluating the $l$-sum
first corresponds to $b=0$. These are just two of an infinite number
of ways to evaluate the double-sum. 

Given that these ambiguous multi-dimensional series are rampant in
this formalism and that frequently the correct method of evaluating
them may be much less obvious,\footnote{For instance, there are cases
  involving triple-sums, where more than one order of evaluating the
  sums give distinct, finite results.} we need a well-motivated
principle that determines the correct way to handle the UV physics.

\subsection{A More Rigorous Derivation}\label{sec:subtleties}

We now present a more rigorous derivation of the intertwining
relations in Equations~\ref{eq:boson-coef}, \eqref{eq:boson-coef-2},
and \eqref{eq:fermion-coef}. By continuously deforming the contour
integral for an initial state mode outward only where the integrand is
holomorphic, we can treat the point $z_0$ more carefully. This
resolves the two problems outlined above.

\begin{figure}[ht]
\subfigure[]
  {\includegraphics[height=3cm, clip=true, trim=0 48 0 48]{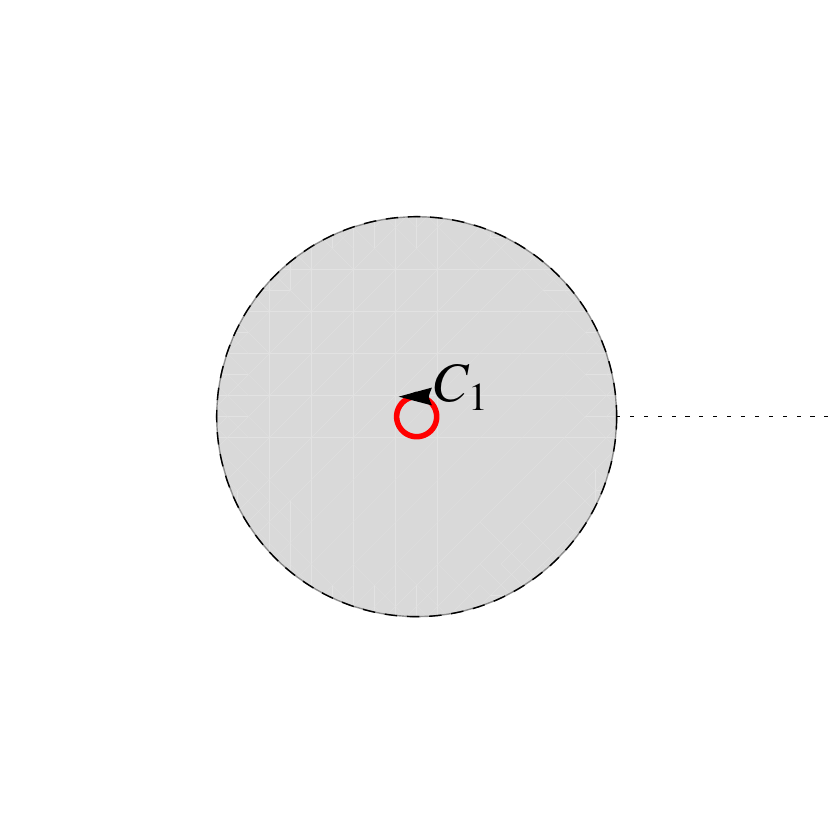}\label{fig:mode1-z}}
\subfigure[]
  {\includegraphics[height=3cm, clip=true, trim=0 48 0 48]{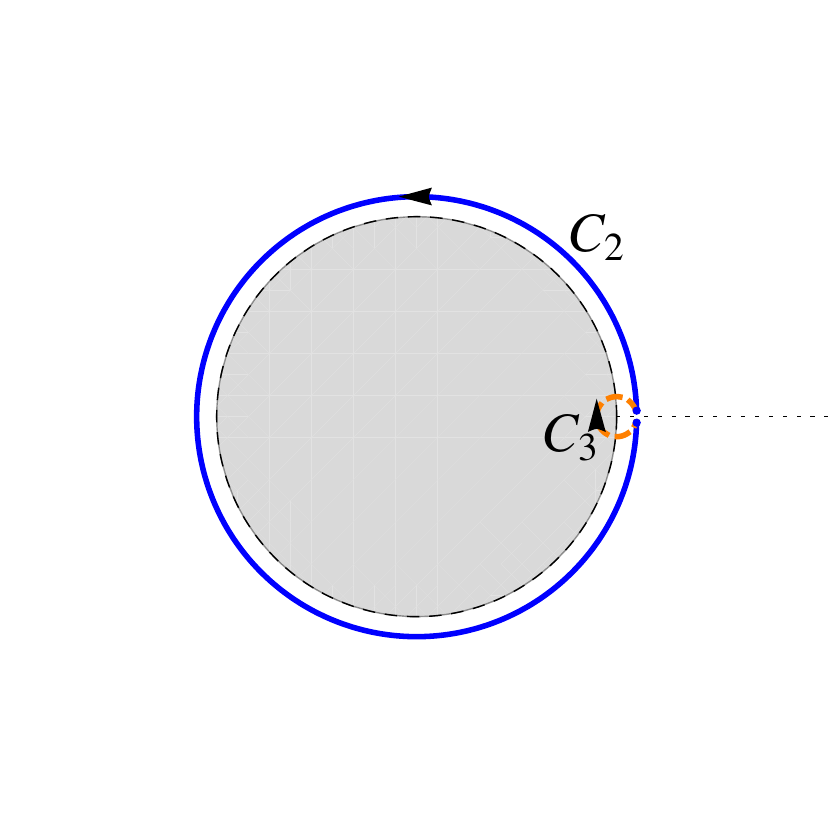}}
  \caption{In the $z$-plane, showing how the contour $C_1$ (solid,
    red) in (a) may be deformed out and around the branch cut into
    contours $C_2$ (solid, blue) and $C_3$ (dashed, orange) in (b).
    The gray circular region is the ``before the twist'' region,
    $|z|<|z_0|$.  The branch cut is indicated by the dashed black line
    extending out from the circle.\label{fig:z-plane-contours}}
\end{figure}

\subsubsection{Bosons}

\begin{figure}[tb]
\begin{center}
\includegraphics[height=3cm]{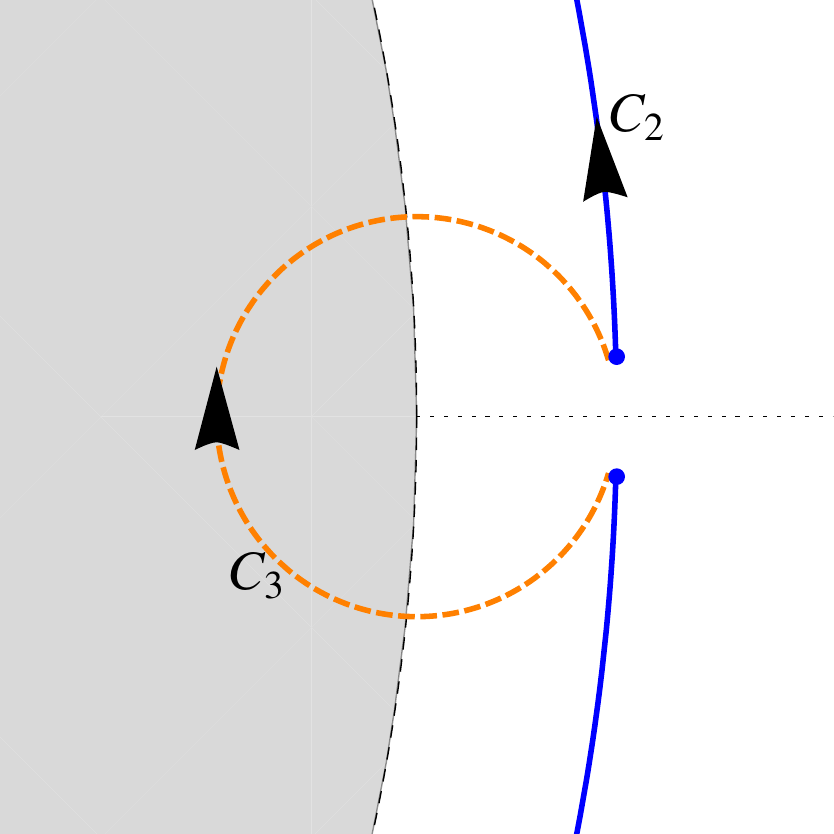}
\caption{A close-up depiction of contours $C_2$ and $C_3$ meeting.
  Note that we have added an artifical gap around the branch cut for
  illustrative purposes only---in fact, both $C_2$ and $C_3$ are full
  circles.\label{fig:close-up}}
\end{center}
\end{figure}

Working with the bosons first, let us note that
\begin{equation}
\alpha_n^{(1)} = \oint_{C_1}\frac{\drm z}{2\pi i}i\pd X^{(1)}(z) z^n,
\end{equation}
where $C_1$ is a circular contour with radius less than $|z_0|$ shown
in Figure~\ref{fig:mode1-z}, and $i\pd X^{(1)}(z)$ is a holomorphic
function except at $z=z_0$ (and excluding any other operator
insertions). Thus, we may deform the contour into an open circle $C_2$
of radius larger than $|z_0|$ and a contour, $C_3$, sneaking around
the branch cut starting at $z=z_0$, as shown in
Figures~\ref{fig:z-plane-contours} and~\ref{fig:close-up}.  We take
$C_3$ to be a circle of radius $\veps$, which we eventually take to
zero. The orientations of the contours are shown in the figures.

We write the above contour integral as
\begin{equation}
\alpha_n^{(1)} = \int_{C_2}\frac{\drm z}{2\pi i}i\pd X^{(1)}(z) z^n 
          + \int_{C_3}\frac{\drm z}{2\pi i}i\pd X^{(1)}(z) z^n.
\end{equation}
The $C_2$ term, is what we have been calculating and is given by (with
$\veps$ corrections)
\begin{calc}
\int_{C_2}\frac{\drm z}{2\pi i} i\pd X(z) z^n &= 
\frac{i}{2}\sum_k\alpha_k \int_{C_2}\frac{\drm z}{2\pi i} z^{n - \frac{k}{2} -1}\\
  &= \frac{1}{4\pi}\sum_k \alpha_k(z_0+\veps)^{n-\frac{k}{2}}
           \int_0^{2\pi}\drm \theta\,e^{i(n-\frac{k}{2})\theta}\\
  &= \frac{1}{2}\alpha_{2n} 
    + \frac{i}{2\pi}\sum_{k\,\text{odd}}\frac{(z_0+\veps)^{n-\frac{k}{2}}}
                                             {n-\frac{k}{2}}\alpha_k.
\end{calc}

\begin{figure}
\subfigure[]
 {\includegraphics[height=6cm]{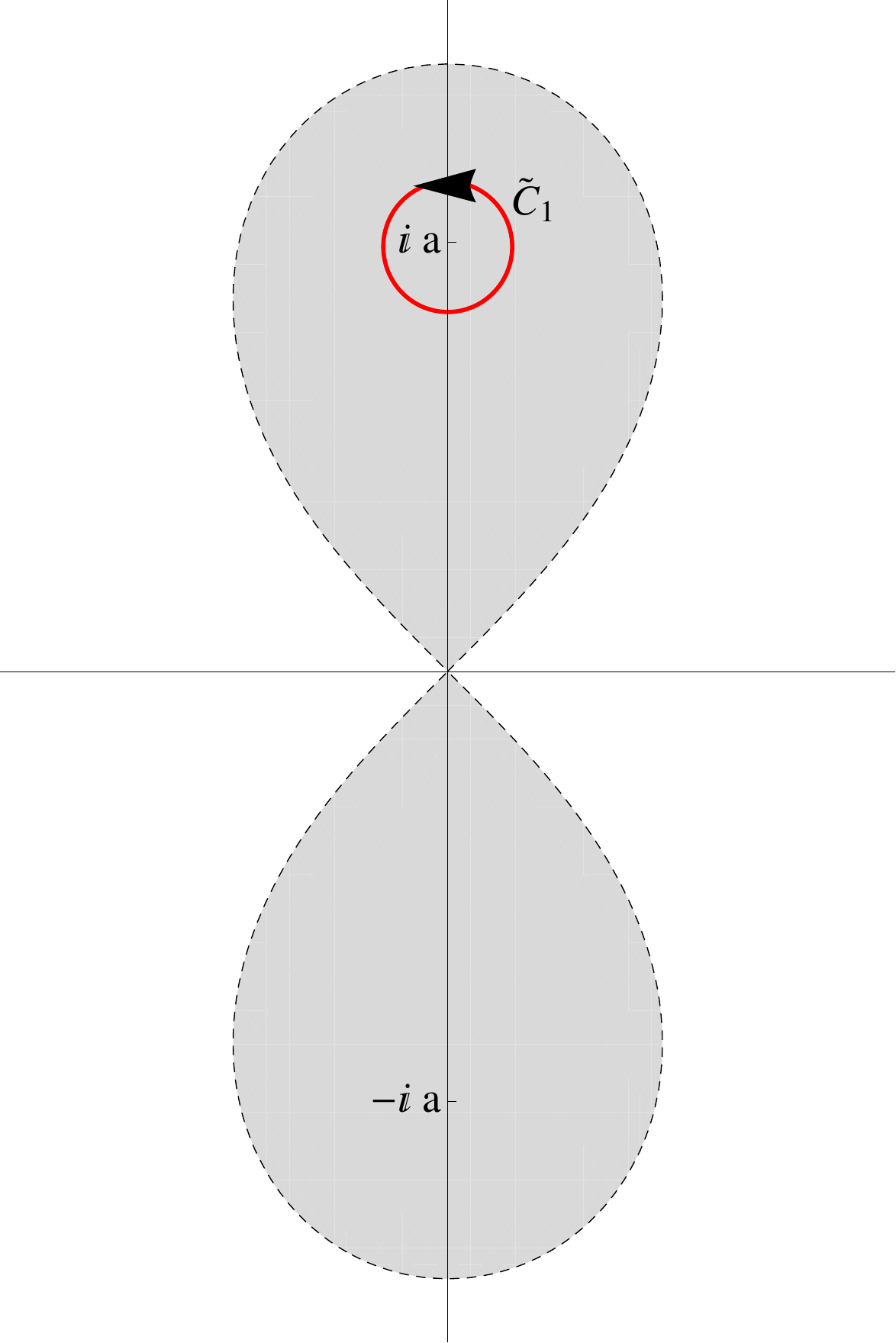}}
\qquad
\subfigure[]
 {\includegraphics[height=6cm]{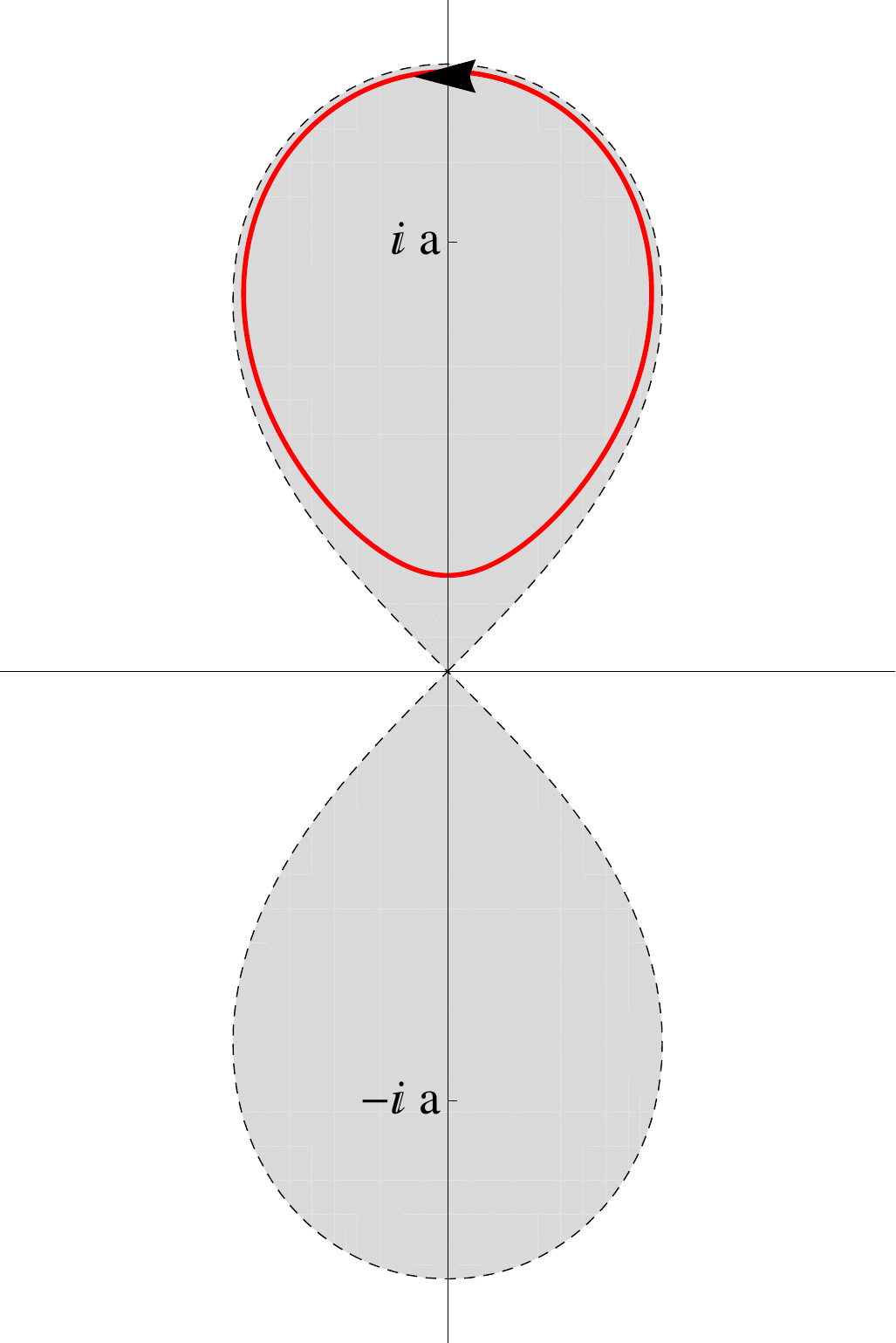}}
\qquad
\subfigure[]
 {\includegraphics[height=6cm]{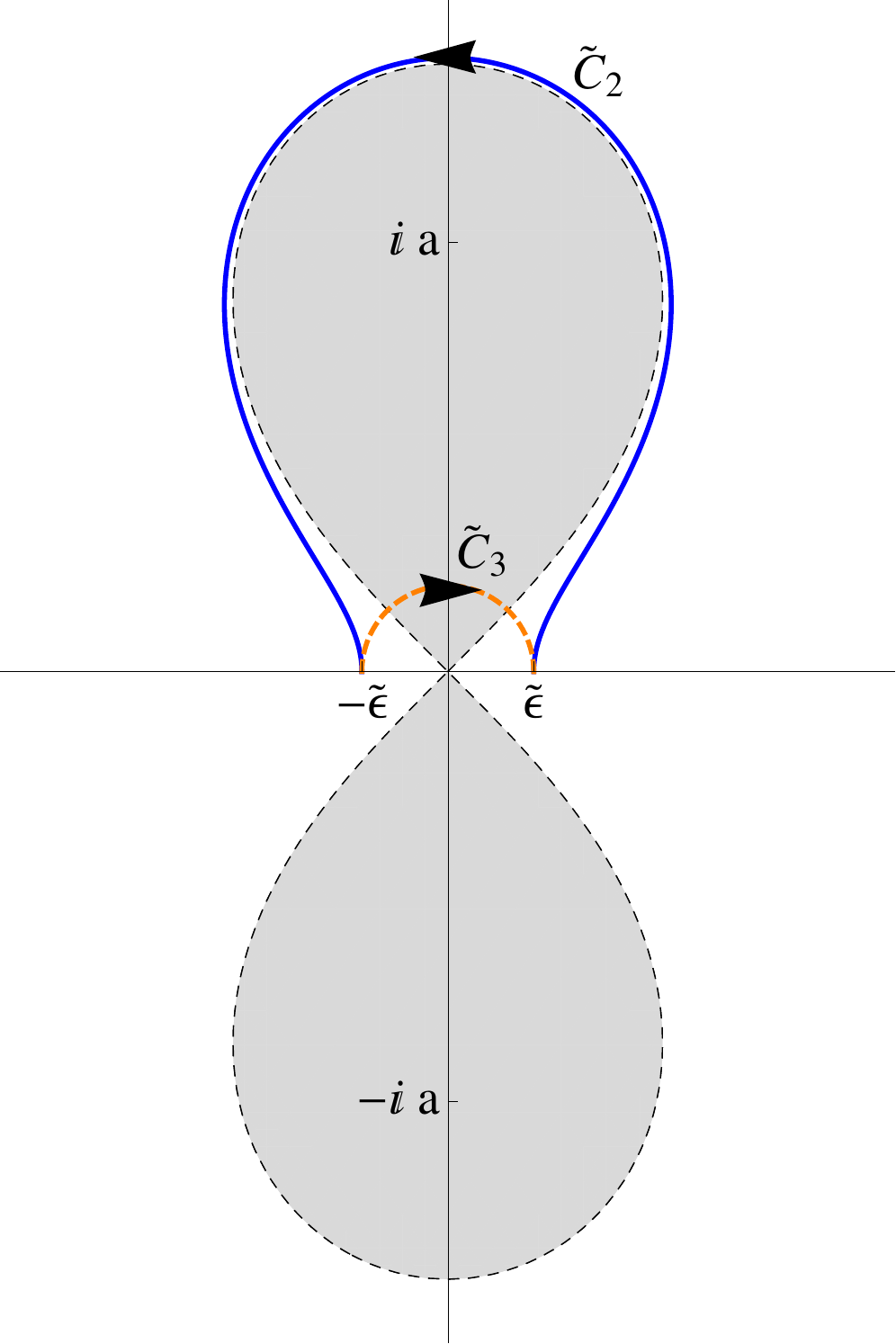}}
 \caption{We show the $t$-plane where there is no branch cut and
   the fields are single-valued.  We start in~(a) with the $\tilde{C}_1$
   contour (solid, red), which we finally deform out into $\tilde{C}_2$ (solid,
   blue) and $\tilde{C}_3$ (dashed, orange) in~(c).
   \label{fig:t-plane-contours}}
\end{figure}

At this point, it becomes necessary to introduce the covering space
where the fields are well-defined. We map to the covering space
coordinate $t$ via
\begin{equation}
z = z_0 + t^2\qquad a^2 = z_0.
\end{equation}
The points $ia$ and $-ia$ are the two images of the origin $z=0$, one
corresponding to each copy of the fields. Mapping the $z$-plane
contours in Figure~\ref{fig:z-plane-contours} to the $t$-plane results
in Figure~\ref{fig:t-plane-contours}.

We compute the $C_3$ term by going to the $t$-plane
\begin{calc}
\int_{C_3}\frac{\drm z}{2\pi i}\pd X(z)z^n = 
  \int_{\tilde{C}_3}\frac{\drm t}{2\pi i} \pd X(t)(z_0 + t^2)^n
  \xrightarrow[\tilde{\veps}\to 0]{} 0,
\end{calc}
where the radius of the semicircular contour in the $t$-plane,
$\tilde{C}_3$, is given by $\tilde{\veps}$. The contour gives zero
contribution since there is no bosonic insertion at $t=0$ and the
integrand is therefore analytic.  As the length of the contour goes to
zero, therefore, so does the integral. Thus, we see that the story for
the bosons is exactly as stated, and in the limit as $\veps\to 0$ we
reproduce our previous result,
\begin{equation}
  \alpha_n^{(1,2)} = \frac{1}{2}\alpha_{2n} 
  \pm \frac{i}{2\pi}\sum_{k\,\text{odd}}
  \frac{z_0^{n-\frac{k}{2}}}{n-\frac{k}{2}}\alpha_k.
\end{equation}

\subsubsection{Fermions}

For the fermions we have
\begin{equation}
\psi^{(1)}_n = \oint_{C_1}\frac{\drm z}{2\pi i}\psi(z) z^{n-\frac{1}{2}},
\end{equation}
which becomes
\begin{equation}
\psi^{(1)}_n = \int_{C_2}\frac{\drm z}{2\pi i}\psi(z) z^{n-\frac{1}{2}}
              +\int_{C_3}\frac{\drm z}{2\pi i}\psi(z) z^{n-\frac{1}{2}}
\end{equation}
The $C_2$ term is what we have computed for the fermions previously,
and is given by
\begin{calc}
\int_{C_2}\frac{\drm z}{2\pi i}\psi(z) z^{n-\frac{1}{2}} &=
   \frac{1}{2}\sum_k\psi_k\int_{C_2}\frac{\drm z}{2\pi i} z^{n-\frac{k}{2}-1}\\
  &= \frac{1}{2}\psi_{2n} 
  \pm \frac{i}{2\pi}\sum_{k\,\text{odd}}\frac{z_0^{n-\frac{k}{2}}}{n-\frac{k}{2}}\psi_k.
\end{calc}
The $C_3$ contribution also goes to zero for the fermions. When one
goes to the $t$-plane, one finds
\begin{equation}
  \int_{\tilde{C}_3}\frac{\drm t}{2\pi i}\psi(t)\sqrt{2t}(z_0 + t^2)^{n-\frac{1}{2}},
\end{equation}
which acts on the spin field $S(t=0)$. The most singular term in the
OPE between $\psi(t)$ and $S(0)$ is proportional to $1/\sqrt{t}$, and
so one again finds that the semi-circular contour vanishes.

At this point, we see the resolution of the first problem we found
with our formal derivation. If the field has an OPE with the
spin-field in the covering space which is singular enough, then the
$C_3$-contribution is nonvanishing in the limit as $\tilde{\veps}\to
0$. This extra contribution gives exactly the correct answer for
$J_0^-$, for instance, as we demonstrate in
Section~\ref{sec:contour-J}.

\subsubsection{Multiple contours}

Having resolved the first problem with the formal derivation, we
should now discuss the UV issues that arise with multiple
contours.\footnote{Recall that the definition of the twist operator
  $\sigma_2$ involves a hole in the $z$-plane, whose radius is
  carefully taken to zero~\cite{lm1}. One might suspect that this
  limit has important consequences for these UV issues and that the
  size of the hole plays the role of a UV cutoff. In fact, the hole is
  taken to zero size \emph{before} any of the issues discussed in this
  paper, and does not play any role here. The actual issue is the
  interaction between neighboring contours, as discussed.}

Let us write $\veps_i$ for the radius of the $i$th mode's $C_3$
semi-circle. For instance, consider an initial state
\begin{equation}
\alpha^{(1)}_{n_1}\alpha^{(1)}_{n_2}\ket{0_R^-}^{(1)}\ket{0_R^-}^{(2)}.
\end{equation}
Then the $C_3$ part coming from $\alpha^{(1)}_{n_1}$ has radius
$\veps_1$ in the $z$-plane and therefore the semi-circle in the
$t$-plane has radius $\sqrt{\veps_1}$. Similarly, for
$\alpha^{(1)}_{n_2}$. If we require that the $C_2$ parts of
$\alpha_{n_1}$ and $\alpha_{n_2}$ preserve the same ordering, then the
semi-circles in the $t$-plane satisfy
\begin{equation}
\veps_2 < \veps_1.
\end{equation}

We now argue that, in fact, we should have $\veps_2 \ll \veps_1$ in
order to use the intertwining relations as we want to. Consider the
two semi-circular $C_3$ contours at leading order in $\veps_1$ and
$\veps_2$:
\begin{calc}
\int_{\tilde{C}_3(\veps_1)}\frac{\drm t}{2\pi i} \pd X(t)(z_0 + t^2)^{n_1}
\int_{\tilde{C}_3(\veps_2)}\frac{\drm t'}{2\pi i} \pd X(t') (z_0 + {t'}^2)^{n_2}
 &\sim \int_{\tilde{C}_3(\veps_1)}\frac{\drm t}{2\pi i} (z_0 + t^2)^{n_1}
\int_{\tilde{C}_3(\veps_2)}\frac{\drm t'}{2\pi i} \frac{ (z_0 + {t'}^2)^{n_2}}{(t-t')^2}\\
 &\sim \int_{\tilde{C}_3(\veps_1)}\frac{\drm t}{2\pi i} z_0^{n_1}
\int_{\tilde{C}_3(\veps_2)}\frac{\drm t'}{2\pi i} \frac{ z_0^{n_2}}{(t-t')^2}\\
 &\sim z_0^{n_1+n_2}\log \frac{1 - \sqrt{\frac{\veps_2}{\veps_1}}}
                              {1+ \sqrt{\frac{\veps_2}{\veps_1}}}\\
 &\sim z_0^{n_1 + n_2} \sqrt{\frac{\veps_2}{\veps_1}} + \cdots.
\end{calc}
This vanishes only if $\veps_2\ll \veps_1$. Note that this argument is
unaffected if the modes are on different copies (the integrals work
out in essentially the same way).

Taking $\veps_2\ll \veps_1$ suggests a particular way to take the
limit for double-sums: if $L_i$ is the cutoff on the
$\alpha_{n_i}^{(1)}$-sum, then we should take
\begin{equation}
L_2 \gg L_1,
\end{equation}
that is, evaluate the $L_2$-sum first and then the $L_1$-sum. This is
exactly the way that we got the correct answer, when the issue was
demonstrated in~\ref{eq:two-boson-problem}.

\subsubsection{The prescription}\label{sec:prescription}

We now develop the precise prescription that resolves the UV
ambiguities. Before we state the prescription, we should mention that
there are two kinds of sums over modes. There are the after-the-twist
intertwining sums, which have ordering ambiguities among themselves,
and there can also be before-the-twist sums on modes before the twist
operator.  For example, consider a composite operator like
\begin{equation}\label{eq:Ln}
J_n^{a(1)} = -\frac{1}{4}{(\sigma^{aT})^\alpha}_\beta\sum_{j =-\infty}^\infty
                \psi^{(1)\dg}_{\alpha A, n-j}\psi^{(1)\beta A}_j.
\end{equation}
In fact, these sums also have UV-limit issues when combined with the
intertwining relations as is demonstrated in
Section~\ref{sec:composite}. If we look at the intertwining relation
in~\ref{eq:boson-coef}, for example, with the implicit cutoff on the
sum,
\begin{equation}\label{eq:boson-coef-cutoff}
\alpha_m^{(2)} = \frac{1}{2}\alpha_{2m} 
                  - \frac{i}{2\pi}\sum_{n\,\text{odd}}^{|n|<L}
                     \frac{z_0^{m-\frac{n}{2}}}{m-\frac{n}{2}}\alpha_n,
\end{equation}
we see that we are approximating a mode $m$ as a linear combination of
modes with UV cutoff $L$. In order for this approximation to become an
exact expansion we must take $L\to\infty$ \emph{with $m$ fixed.} That
is we need to have much higher frequency modes in our sum than the
mode that we are expanding. Therefore, we need to cutoff the
before-the-twist sum in~\eqref{eq:Ln} and ensure that its cutoff is
much less than the after-the-twist cutoff
in~\eqref{eq:boson-coef-cutoff}.

Before proceeding, let us consider more generally what sort of
multi-dimensional series we can get in this formalism. If we have a
bunch of modes in the initial state that we pull across using the
intertwining relations, then there are several different types of
terms that can arise.  There is a term in which all of the positive
modes act on $\ket{\chi}$ separately, and we are left with a product
of one-dimensional sums that result in either~\eqref{eq:single-boson}
or~\eqref{eq:single-fermion}. Then, there are terms where the positive
modes from one sum contract with negative modes of another. This gives
a double sum, like the one in~\ref{eq:ambig-sum}. If there is a sum
on the before-the-twist modes, then we can get a triple sum if those
two modes contract. This is the most complicated sum possible.

Finally, we are ready to state the prescription that ensures that the
multi-dimensional series converge to the correct answer. The
prescription is
\begin{enumerate}
\item The after-the-twist intertwining sums should be performed from
  innermost contour (right-most mode) to outermost contour (left-most
  mode). This ensures that the $C_3$-terms can be dropped.
\item Any sums on before-the-twist modes should be performed
  \emph{last}. There is no UV-ambiguity among the
  before-the-twist sums.
\end{enumerate}

Note that with this prescription we have a weakened version of our
goal. Because of the UV sensitive series, one cannot directly map
operators to operators since one requires knowledge of what other
modes are around in order to correctly evaluate the series.  While the
modes that we start with before the twist operator may commute, we
need to think about them pulling across the twist operator in a
particular order. It is in this sense, that we have intertwining
relations and not Bogolyubov coefficients.

\section{An example: intertwining relations for $J^a_n$}\label{sec:example}

There are two equivalent ways of calculating the effect of the twist
operator on composite operators such as $J^a_n$. One way is to use the
contour deformation method described in this paper, being careful not
to throw away the small contour $C_3$. The other way is to write $J^a$
as the product of fermion modes and use the intertwining relations for
the fermions, being careful to use the prescriptions described in
Section~\ref{sec:prescription}.

\subsection{The contour method}\label{sec:contour-J}

We first describe the contour method mentioned above. If we go through
the argument described in Section~\ref{sec:subtleties} then
\begin{equation}
J^{a(1)}_n = \frac{1}{2}J^a_{2n} + \frac{i}{2\pi}\sum_{k\,\text{odd}}
        \frac{z_0^{n-\frac{k}{2}}}{n-\frac{k}{2}}J^a_k
        + \lim_{\veps\to 0}\int_{C_3}\frac{\drm z}{2\pi i} J^{a(1)}(z)z^n, 
\end{equation}
but this time we find a nonvanishing contribution coming from the
$C_3$ contour. As we pull the $J^{a(1)}_n$ contour out, it acts on the
twist operator (or equivalently, the spin field in the covering space)
and can switch a $\sigma_2^+$ to $\sigma_2^-$. Therefore, let us
consider $\sigma_2^{\alpha}$. 

We can evaluate the $C_3$ contour by going to the $t$-plane, where
$\sigma_2^\alpha(z_0)$ leaves only a spin field $S^\alpha(t=0)$. The
image of $C_3$ in the $t$-plane, $\tilde{C}_3$, is a semi-circle
around the origin as shown in Figure~\ref{fig:t-plane-contours}. Thus,
the $\tilde{C}_3$ contour implicitly acts on the spin field:
\begin{calc}
\int_{C_3}\frac{\drm z}{2\pi i} J^{(1)a}(z)z^n
 &= \left[\int_{\tilde{C}_3}\frac{\drm t}{2\pi i} J^a(t)(z_0 + t^2)^n\right]S^\alpha(0)\\
 &= \frac{1}{2}{(\sigma^{aT})^\alpha}_\beta S^\beta(0)
    \int_{\tilde{C}_3}\frac{\drm t}{2\pi i}\frac{(z_0 + t^2)^n}{t}\\
 &= \frac{1}{2}{(\sigma^{aT})^\alpha}_\beta S^\beta(0)
     \frac{1}{2\pi i}\left(z_0^n \log t \bigg|^{\veps}_{-\veps} + \bigO(\veps)\right)\\
 &=  -\frac{z_0^n}{4}{(\sigma^{aT})^\alpha}_\beta S^\beta(0),
\end{calc}
where we have given the result after taking $\veps \to 0$. When we go
back to the $z$-plane we should write the above result as
\begin{equation}
\sigma_2^\alpha(z_0)J^{a(1)}_n  = \left[\frac{1}{2}J^a_{2n} 
          +\frac{i}{2\pi}\sum_{k\,\text{odd}}\frac{z_0^{n-\frac{k}{2}}}{n-\frac{k}{2}}J^a_k\right]
            \sigma_2^\alpha(z_0)
             - \frac{z_0^n}{4}{(\sigma^{aT})^\alpha}_\beta\sigma_2^\beta(z_0).
\end{equation}
If we had considered $J_n^{a(2)}$, then we obtain a similar result. We
can summarize the two relations and write them in a suggestive form as
\begin{equation}\label{eq:J-intertwining}
\sigma_2^\alpha(z_0)J^{a(1,2)}_n  = \left[\frac{1}{2}J^a_{2n} 
          \pm\frac{i}{2\pi}\sum_{k\,\text{odd}}\frac{z_0^{n-\frac{k}{2}}}{n-\frac{k}{2}}J^a_k\right]
            \sigma_2^\alpha(z_0)
             - \frac{z_0^n}{2}\com{J^a_0}{\sigma_2^\alpha(z_0)}. 
\end{equation}
We see that if we use these intertwining relations, then we get the
correct answer in~\ref{eq:wrong-J-com}.

While the above intertwining relation is correct, it may not be the
most useful form. Because we have switched the charge on the twist
operator, we now must think about a new state $\ket{\chi}$ created by
the negatively-charged operator acting on the vacuum.

\subsubsection{An example}

For concreteness, let us consider
\begin{equation}
\sigma_2^+(z_0) J_n^{-(1)}\ket{0^-}^{(1)}\ket{0^-}^{(2)},\qquad n<0,
\end{equation}
then we have
\begin{equation}
\left[\frac{1}{2}J^-_{2n} + \frac{i}{2\pi}
         \sum_{k\,\text{odd}}\frac{z_0^{n-\frac{k}{2}}}{n-\frac{k}{2}}J^-_k
- \frac{z_0^n}{2}J_0^-\right]\ket{\chi}.
\end{equation}
Thus, our first task is to compute the three terms,
\begin{equation}
J_{2n}^-\ket{\chi}\qquad
J_k^-\ket{\chi}\,k\,\text{odd}\qquad 
J_0^-\ket{\chi}.
\end{equation}

There are no real complications in working the terms out. For
instance, for the first term, one starts with\footnote{Note that the
  $J$ after the twist has an extra factor of $1/2$ from before the
  twist, which arises from the fermion--fermion anticommutator having
  an extra factor of $2$ after the twist.}
\begin{equation}
J^-_{2n} = -\frac{1}{4}\sum_j\psi^\dg_{+A, 2n - j}\psi^{-A}_{j},
\end{equation}
then breaks the sum into terms with both modes negative and terms with
odd positive modes that act on $\ket{\chi}$. One can write the result
in the form
\begin{equation}
J^-_{2n}\ket{\chi} = 
-\frac{1}{4}\sum_{2n+1\leq j \leq -1} \psi^\dg_{+A, 2n-j}\psi^{-A}_j\ket{\chi}
 + \sum_{j,p\,\text{odd}^+}\gamma^F_{jp}\psi^\dg_{+A, -p}\psi^{-A}_{2n-j}\ket{\chi}.
\end{equation}
Similarly,  $J^-_0$ becomes
\begin{equation}
J_0^-\ket{\chi} = \sum_{j,p\,\text{odd}^+}\gamma^F_{jp}\psi^\dg_{+A, -j}\psi^{-A}_{-p}\ket{\chi}.
\end{equation}

The $J_k^-$ term is not much work, but there are two distinct cases,
corresponding to whether or not there are terms where both $\psi$s are
raising operators: $k\leq -3$ and $k\geq -1$. One finds
\begin{equation}
J^-_k\ket{\chi} = -\frac{1}{4}\sum_{k+1\leq j \leq -1} \psi^\dg_{+A, k-j}\psi^{-A}_j\ket{\chi}
         + \sum_{j,p\,\text{odd}^+} \gamma^F_{jp}\psi^\dg_{+A, k-j}\psi^{-A}_{-p}\ket{\chi}
         \qquad k\leq -3, \text{odd}
\end{equation}
and
\begin{equation}
J^-_k\ket{\chi} = \sum_{j,p\,\text{odd}^+}\gamma^F_{k+j+1, p}\psi^\dg_{+A, -j-1}\psi^{-A}_{-p}
                 \ket{\chi}
\qquad k\geq -1,\text{odd}.
\end{equation}
The slightly more difficult task is the sum over $k$, which can be written as
\begin{calc}
\sum_{k\,\text{odd}}\frac{z_0^{n-\frac{k}{2}}}{n-\frac{k}{2}}J^-_k\ket{\chi}
 &= \sum_{k\,\text{odd}}^{k\leq -3}\frac{z_0^{n-\frac{k}{2}}}{n-\frac{k}{2}}J^-_k\ket{\chi}
   +\sum_{k\,\text{odd}}^{k\geq -1}\frac{z_0^{n-\frac{k}{2}}}{n-\frac{k}{2}}J^-_k\ket{\chi}\\
 &= 
-\frac{1}{2}\sum_{j,p\,\text{odd}^+}\frac{z_0^{n+\frac{j+p+1}{2}}}{n+\frac{j+p+1}{2}}
       \psi^\dg_{+A, -j-1}\psi^{-A}_{-p}\\
&\qquad+\sum_{j\,\text{odd}}^{j\geq 3}\sum_{p\,\text{odd}^+}
     \left(
       \sum_{k\,\text{odd}}^{-j\leq k\leq -3}\frac{z_0^{n-\frac{k}{2}}}{n-\frac{k}{2}}
            \gamma^F_{k+j+1,p}
     \right)\psi^\dg_{+A, -j-1}\psi^{-A}_{-p}\\
&\qquad+\sum_{j,p\,\text{odd}^+}
       \left(
         \sum_{k\,\text{odd}}^{k\geq -1}\frac{z_0^{n-\frac{k}{2}}}{n-\frac{k}{2}}\gamma^F_{k+j+1,p}
       \right)\psi^\dg_{+A, -j-1}\psi^{-A}_{-p},
\end{calc}
after a few manipulations. Note that all of the terms on the right
implicitly act on $\ket{\chi}$. Finally, using~\eqref{eq:gammaF-sum-1}
one arrives at
\begin{calc}
\sum_{k\,\text{odd}}\frac{z_0^{n-\frac{k}{2}}}{n-\frac{k}{2}}J^-_k\ket{\chi}
 =-\sum_{j,p\,\text{odd}^+}
    \frac{z_0^{n+\frac{j+p+1}{2}}}{2(n+\frac{j+p+1}{2})}
     \frac{\Gamma(\frac{p}{2})\Gamma(-n-\frac{j}{2})}
                {\Gamma(\frac{p+1}{2})\Gamma(-n-\frac{j+1}{2})}
              \psi^\dg_{+A,-j-1}\psi^{-A}_{-p}\ket{\chi}
\end{calc}

Putting it all together, we have
\begin{equation}\begin{split}\label{eq:J-on-vac}
\sigma_2^+(z_0) J_n^{-(1)}\ket{0^-}^{(1)}\ket{0^-}^{(2)}
 &= -\frac{1}{8}\sum_{j=n+1}^{-1}\psi^\dg_{+A, 2n-2j}\psi^{-A}_{2j}\ket{\chi}\\
&\quad-\frac{i}{4\pi}
\sum_{j=1}^\infty\sum_{p\,\text{odd}^+}
    \frac{z_0^{n+j+\frac{p}{2}}}{n+j+\frac{p}{2}}
     \frac{\Gamma(\frac{p}{2})\Gamma(-n-j+\frac{1}{2})}
                {\Gamma(\frac{p+1}{2})\Gamma(-n-j)}
              \psi^\dg_{+A,-2j}\psi^{-A}_{-p}\ket{\chi}\\
&\quad+\frac{1}{2}\sum_{j,p\,\text{odd}^+}
\left[-\frac{1}{4}\delta_{j+p+2n,0} + \frac{\gamma^F_{2n+p,j}+\gamma^F_{2n+j,p}}{2}
      - z_0^n \frac{\gamma^F_{jp} + \gamma^F_{pj}}{2}\right]\psi^\dg_{+A, -j}\psi^{-A}_{-p}
\ket{\chi},
\end{split}\end{equation}
where for ease of comparison we have broken the result into
even--even, even--odd, and odd--odd terms. In the above expression, we
define $\gamma^F$ with negative indices to be zero. Furthermore, we
have explicitly symmetrized over $j$ and $p$ in the last term since
\begin{equation}
\psi^\dg_{+A, -j}\psi^{-A}_{-p} = \psi^\dg_{+A, -p}\psi^{-A}_{-j}.
\end{equation}
Below, we compare this result with what one finds when one breaks the
$J^{-(1)}_{n}$ into fermions and uses the fermion intertwining
relations.

\subsection{The composite method}\label{sec:composite}

We now show how to reproduce the above result by using the fermion
intertwining relations and our prescription. 

We start by writing
\begin{equation}
J^{a(1)}_n = -\frac{1}{4}{(\sigma^{aT})^\alpha}_\beta\sum_{j =-\infty}^\infty
                \psi^{(1)\dg}_{\alpha A, n-j}\psi^{(1)\beta A}_j,
\end{equation}
and note that the sum on $j$ is a sum over ``before-the-twist'' modes
and therefore should be evaluated last according to the prescription.
We then can use our intertwining relations to write this directly as
\begin{equation}\label{eq:J-sum}
J^{a(1)}_n = -\frac{1}{4}{(\sigma^{aT})^\alpha}_\beta\sum_{j =-\infty}^\infty
  \left[\frac{1}{2}\psi^\dg_{\alpha A, 2n-2j} 
           + \frac{i}{2\pi}\sum_{k\,\text{odd}}\frac{z_0^{n-j-\frac{k}{2}}}{n-j-\frac{k}{2}}
                     \psi^\dg_{\alpha A, k}\right]
  \left[\frac{1}{2}\psi^{\beta A}_{2j} + \frac{i}{2\pi}\sum_{l\,\text{odd}}
                             \frac{z_0^{j-\frac{l}{2}}}{j-\frac{l}{2}}\psi_l^{\beta A}\right].
\end{equation}
At this point, we can say nothing further until we know what
$J^{a(1)}_n$ acts on. Of course, when faced with an expression like
the above it is rather tempting to evaluate the $j$-sum using
\begin{equation}
\sum_{j=-\infty}^\infty \frac{1}{(n-j-\frac{k}{2})(j-\frac{l}{2})} = 
   -\pi^2\delta_{\frac{l}{2}, n-\frac{k}{2}} \qquad k,l\,\text{odd},
\end{equation}
which immediately leads to the false relation
\begin{equation}
J^{a(1)}_n = \frac{1}{2}J^a_{2n} +
  \frac{i}{2\pi}\sum_{k\,\text{odd}}\frac{z_0^{n-\frac{k}{2}}}{n-\frac{k}{2}}
  J^a_k
\qquad \text{(false relation)}.
\end{equation}
This demonstrates the need for the restriction on before-the-twist
sums.

In order to proceed and compare to~\eqref{eq:J-intertwining}, let us
again consider the state
\begin{equation}\label{eq:J-state}
\sigma_2^+(z_0)J^{-(1)}_n\ket{0_R^-}^{(1)}\ket{0_R^-}^{(2)}.
\end{equation}
Now, one could make the $j$-sum in~\eqref{eq:J-sum} finite by
considering the above; however, our prescription ensures that one gets
the correct answer even if one leaves it as an infinite series.

If we act on the vacuum as in~\eqref{eq:J-state}, then we
get~\eqref{eq:J-sum} acting on $\ket{\chi}$. The $\psi^{\beta A}$ is
the rightmost mode and so the $l$-sum should be evaluated first,
followed by the $k$-sum, and then finally the $j$-sum. We can
use~\eqref{eq:single-fermion} to quickly read off the result of the
$l$-sum. There are now two distinct terms to consider for the $k$-sum.
There is the possibility of the $\psi^\dg_{\alpha A}$ contracting with
the result of the $l$-sum, and the $\psi^\dg$ can pass through and act
on $\ket{\chi}$ (and we can again use~\eqref{eq:single-fermion}). The
contraction gives zero. For other composite operators, however, the
contraction term can be nonzero.

Following the above procedure, we get
\begin{equation}\begin{split}
-\frac{1}{2}\sum_{j=-\infty}^\infty
  &\left[\frac{1}{2}\psi^\dg_{+A, 2(n-j)} + \frac{i}{2\pi}\sum_{p\,\text{odd}^+}
    \frac{z_0^{n-j+\frac{p}{2}}}{n-j+\frac{p}{2}}
    \frac{\Gamma(\frac{p}{2})\Gamma(-n+j+\frac{1}{2})}{\Gamma(\frac{p+1}{2})\Gamma(-n+j)}
    \psi^\dg_{+A, -p}\right]\\
  &\left[\frac{1}{2}\psi^{-A}_{2j} + \frac{i}{2\pi}\sum_{q\,\text{odd}^+}
    \frac{z_0^{j+\frac{q}{2}}}{j+\frac{q}{2}}
    \frac{\Gamma(\frac{q}{2})\Gamma(-j+\frac{1}{2})}{\Gamma(\frac{q+1}{2})\Gamma(-j)}
    \psi^{-A}_{-q}\right].
\end{split}\end{equation}
There are four terms from the above multiplication. 

The even--even term is
\begin{equation}
(\text{even--even})=
-\frac{1}{8}\sum_{j=n+1}^{-1}\psi^\dg_{+A, 2n-2j}\psi^{-A}_{2j}.
\end{equation}
One can show that the two even--odd cross-terms are identical, and sum
to
\begin{equation}
(\text{even--odd}) =
-\frac{i}{4\pi}\sum_{j=-\infty}^{-1}
\sum_{p\,\text{odd}^+}\frac{z_0^{n-j+\frac{p}{2}}}{n-j+\frac{p}{2}}
    \frac{\Gamma(\frac{p}{2})\Gamma(-n+j+\frac{1}{2})}{\Gamma(\frac{p+1}{2})\Gamma(-n+j)}
    \psi^\dg_{+A, -p}\psi^{-A}_{2j}
\end{equation}
One may write the odd--odd term as
\begin{equation}
(\text{odd--odd}) = \frac{1}{8\pi^2}\sum_{p,q\,\text{odd}^+}\psi^\dg_{+A, -p}\psi^{-A}_{-q}
        z_0^{n+\frac{p+q}{2}}
        \frac{\Gamma(\frac{p}{2})\Gamma(\frac{q}{2})}{\Gamma(\frac{p+1}{2})\Gamma(\frac{q+1}{2})}
        S(n,p,q),
\end{equation}
where
\begin{equation}
S(n,p,q) = \sum_{j=n+1}^{-1}\frac{1}{(n-j+\frac{p}{2})(j+\frac{q}{2})}
                       \frac{\Gamma(-n+j+\frac{1}{2})\Gamma(-j+\frac{1}{2})}
               {\Gamma(-n+j)\Gamma(-j)}.
\end{equation}

Comparing the above to the three terms in Equation~\eqref{eq:J-on-vac},
one finds agreement provided
\begin{equation}
\frac{z_0^{n+\frac{j+p}{2}}}{8\pi^2}\frac{\Gamma(\frac{j}{2})\Gamma(\frac{p}{2})}
                                          {\Gamma(\frac{j+1}{2})\Gamma(\frac{p+1}{2})}
                                          S(n,j,p)
 = \frac{1}{2}\left[-\frac{1}{4}\delta_{j+p+2n,0} + \frac{\gamma^F_{2n+p,j}+\gamma^F_{2n+j,p}}{2}
      - z_0^n \frac{\gamma^F_{jp} + \gamma^F_{pj}}{2}\right].
\end{equation}
This equation follows from the identity
\begin{multline}\label{eq:hard-id}
\sum_{k=0}^\mu\frac{1}{(\mu-k-\frac{\alpha}{2})(k-\frac{\beta}{2})}
   \frac{\Gamma(\mu -k + \frac{3}{2})\Gamma(k+\frac{3}{2})}
        {\Gamma(\mu - k + 1)\Gamma(k+1)}\\
=\pi - \frac{\pi}{\frac{\alpha + \beta}{2} -\mu}
\left(\frac{\Gamma(\frac{\alpha+3}{2})\Gamma(\frac{\alpha}{2}-\mu)}
           {\Gamma(\frac{\alpha}{2}+1)\Gamma(\frac{\alpha-1}{2}-\mu)}
+\frac{\Gamma(\frac{\beta+3}{2})\Gamma(\frac{\beta}{2}-\mu)}
           {\Gamma(\frac{\beta}{2}+1)\Gamma(\frac{\beta-1}{2}-\mu)}
\right),
\end{multline}
where the left-hand side is $S(n,p,q)$ with $\mu = -n-2$, $\alpha= p
-2$, and $\beta = q-2$. If we call the above sum $F(\mu, \alpha,
\beta)$, then one can prove the identity by showing that both sides of
Equation~\eqref{eq:hard-id} obey the four-term recursion relation
\begin{multline}
  f_0(\mu,\alpha, \beta) F(\mu, \alpha, \beta) 
+ f_1(\mu,\alpha, \beta) F(\mu+1, \alpha, \beta)\\
+ f_2(\mu,\alpha, \beta) F(\mu+2, \alpha, \beta) 
+ f_3(\mu,\alpha, \beta) F(\mu+3, \alpha, \beta) 
 = 0,
\end{multline}
with
\begin{subequations}
\begin{align}
f_0(\mu, \alpha,\beta) &= (-3+\alpha -2 \mu ) (\alpha +\beta -2 \mu ) (3-\beta +2 \mu ) (\alpha +\beta -6 (3+\mu ))\\
f_1(\mu, \alpha,\beta) &= 3 (\alpha +\beta -2 (1+\mu )) \big(-2 \beta ^2 (2+\mu )+\alpha ^2 (\beta -2 (2+\mu ))+\beta  (79+8 \mu  (9+2 \mu ))\notag\\
&+\alpha  \left(79+\beta ^2+8 \mu  (9+2 \mu )-2 \beta  (12+5 \mu )\right)-2 (128+\mu  (175+4 \mu  (20+3 \mu )))\big)\\
f_2(\mu, \alpha,\beta) &= 3 (\alpha +\beta -2 (2+\mu )) \big(\beta ^2 (5+2 \mu )+\alpha ^2 (5-\beta +2 \mu )+4 (2+\mu ) \left(43+32 \mu +6 \mu ^2\right)\notag\\
&\quad-2 \beta  (47+\mu  (39+8 \mu ))-\alpha  \left(94+\beta ^2-2 \beta  (12+5 \mu )+2 \mu  (39+8 \mu )\right)\big)\\
f_3(\mu, \alpha,\beta) &= (\alpha +\beta -6 (2+\mu )) (\alpha -2 (3+\mu )) (\beta -2 (3+\mu )) (\alpha +\beta -2 (3+\mu )).
\end{align}
\end{subequations}
Thus the identity holds by induction once one confirms that it holds
for $\mu=0,1,2$. It may be helpful to use an algebraic manipulation
program such as \textit{Mathematica} to show the above.

\section{Conclusion}\label{sec:conclusion}

We presented a method for computing the effect of the blow-up mode
deformation operator on untwisted states of the D1D5 CFT. An
alternative method was developed in~\cite{acm3}; however, the method
developed here comes in the form of intertwining relations for the
2-twist operator, $\sigma_2^+(z_0)$. This method, therefore, is more
direct and of theoretical interest since it makes the physics more
transparent. We would be remiss, if we did not note that both the
method in~\cite{acm3} and that presented here should generalize to
higher order twist operators.

Using the technology and understanding developed in~\cite{acm2, acm3}
and here, we hope to address some important outstanding questions
concerning the D1D5 CFT and black hole physics. In particular, we hope
to elucidate the nature of the proposed
``non-renormalization'' theorem~\cite{dmw02}; how states of the CFT fragment,
thereby ``scrambling'' information~\cite{susskind,preskill}; black
hole formation; and the in-falling observer. All of these are
important and deep issues that will probably require lots of work to
resolve; however, this paper should serve as a step toward these
goals. All of the discussion here and in~\cite{acm2, acm3} has been
off-shell. A more modest next step would be to analyze the physics
that arise on-shell. Another task is to systematically analyze the
combinatorial factors that arise at different orders in perturbation
theory, perhaps using the technology developed in~\cite{rastelli1,
  rastelli2}.

\section*{Acknowledgments}

We are grateful for insightful conversations with and sundry
contributions from S.~Mathur.  We thank K.~Bobkov, B.~Dundee,
M.~Randeria, M.~Shigemori, and Y.~K.~Srivastava for several helpful
conversations.  The work of SGA is supported in part by DOE grant
DE-FG02-91ER-40690.  The work of BDC is supported by the Foundation
for Fundamental Research on Matter.

\appendix

\section{Notation and the CFT algebra} \label{ap:CFT-notation}

For reference, we give the notation and conventions we use (see
also~\cite{acm2,acm3}). What we describe below is the field content
and algebra of a single copy of the CFT in the untwisted sector.

We have 4 real left-moving fermions, $\psi_1$ through $\psi_4$, which
we group into doublets $\psi^{\alpha A}$:
\begin{subequations}
\begin{align}
\begin{pmatrix}\psi^{++}\\ \psi^{-+}\end{pmatrix} &= \frac{1}{\sqrt{2}}
\begin{pmatrix}\psi_1+i\psi_2\\ 
               \psi_3+i\psi_4
\end{pmatrix}\\
\begin{pmatrix}\psi^{+-}\\ \psi^{--}\end{pmatrix} &=\frac{1}{\sqrt{2}}
\begin{pmatrix}\psi_3-i\psi_4\\
              -(\psi_1-i\psi_2)
\end{pmatrix}
\end{align}
\end{subequations}
Here $\alpha=(+,-)$ is an index of the subgroup $SU(2)_L$ of rotations
on $S^3$ and $A=(+,-)$ is an index of the subgroup $SU(2)_1$ from
rotations in $T^4$. The 2-point functions are
\begin{equation}
\langle\psi^{\alpha A}(z)\psi^{\beta B}(w)\rangle=-\epsilon^{\alpha\beta}\epsilon^{AB}{1\over z-w},
\end{equation}
where we have defined the $\epsilon$ symbol as
\begin{equation}
\epsilon_{+-}=1\qquad\epsilon^{+-}=-1.
\end{equation}
Hermitian conjugation is defined as
\begin{equation}
\psi^\dg_{\alpha A} = -\epsilon_{\alpha\beta}\epsilon_{A B}\psi^{\beta B}.
\end{equation}

There are 4 real left-moving bosons, $X_1$ through $X_4$, which can be
grouped into a matrix
\begin{equation}
X_{A\dot A}= \frac{1}{\sqrt{2}} X_i \sigma_i=\frac{1}{\sqrt{2}} \begin{pmatrix} X_3+iX_4& X_1-iX_2\\ X_1+iX_2&-X_3+iX_4\end{pmatrix},
\end{equation}
where $\sigma_i$ with $i=1,2,3$ are the Pauli matrices and $\sigma_4 =
iI$. The 2-point functions are
\begin{equation}
\langle\p X_{A\dot A}(z) \p X_{B\dot B}(w)\rangle={1\over (z-w)^2}\epsilon_{AB}\epsilon_{\dot A\dot B}.
\end{equation}

The bosonic  and fermionic modes satisfy
\begin{equation}
\com{\alpha_{m,A\dot{A}}}{\alpha_{n,B\dot{B}}} = m\epsilon_{AB}\epsilon_{\dot{A}\dot{B}}\delta_{m+n,0},
\end{equation}
and
\begin{equation}
\ac{\psi^{\alpha A}_m}{\psi^{\beta B}_n} = -\epsilon^{\alpha\beta}\epsilon^{AB}\delta_{m+n,0},
\end{equation}
respectively.

The chiral algebra is generated by the operators
\begin{subequations}
\begin{align}
J^a &=-{1\over 4}(\psi^\dagger)_{\alpha A} (\sigma^{Ta})^\alpha{}_\beta \psi^{\beta A}\\
G^\alpha_{\dot A} &= \psi^{\alpha A} \p X_{A\dot A}, ~~~(G^\dagger)_{\alpha}^{\dot A}=(\psi^\dagger)_{\alpha A} \p (X^\dagger)^{A\dot A}\\
T &=-{1\over 2} (\p X^\dagger)^{A\dot A}\p X_{A\dot A}-{1\over 2} (\psi^\dagger)_{\alpha A} \p \psi^{\alpha A}\\
(G^\dagger)_{\alpha}^{\dot A} &=-\epsilon_{\alpha\beta} \epsilon^{\dot A\dot B}G^\beta_{\dot B}, ~~~~G^{\alpha}_{\dot A}=-\epsilon^{\alpha\beta} \epsilon_{\dot A\dot B}(G^\dagger)_\beta^{\dot B}.
\end{align}\end{subequations}
These operators generate the algebra
\begin{subequations}
\begin{align}
J^a(z) J^b(w) &\sim \delta^{ab} {\h\over (z-w)^2}+i\epsilon^{abc} {J^c\over z-w}\\
J^a(z) G^\alpha_{\dot A} (z') &\sim {1\over (z-z')}\h (\sigma^{aT})^\alpha{}_\beta G^\beta_{\dot A}\\
G^\alpha_{\dot A}(z) (G^\dagger)^{\dot B}_\beta(z') &\sim 
      -{2\over (z-z')^3}\delta^\alpha_\beta \delta^{\dot B}_{\dot A}
      - \delta^{\dot B}_{\dot A}  (\sigma^{Ta})^\alpha{}_\beta 
              [{2J^a\over (z-z')^2}+{\p J^a\over (z-z')}]
      -{1\over (z-w)}\delta^\alpha_\beta \delta^{\dot B}_{\dot A}T\\
T(z)T(z') &\sim {3\over (z-z')^4}+{2T\over (z-z')^2}+{\p T\over (z-z')}\\
T(z) J^a(z') &\sim {J^a\over (z-z')^2}+{\p J^a\over (z-z')}\\
T(z) G^\alpha_{\dot A}(z') &\sim {{3\over 2}G^\alpha_{\dot A}\over (z-z')^2}  + {\p G^\alpha_{\dot A}\over (z-z')}. 
\end{align}
\end{subequations}

Note that
\begin{equation}
J^a(z) \psi^{\gamma C}(w)\sim {1\over 2} {1\over z-w} (\sigma^{aT})^\gamma{}_\beta \psi^{\beta C}.
\end{equation}

The above OPE algebra gives the commutation relations
\begin{subequations}
\begin{eqnarray}
\com{J^a_m}{J^b_n} &=& \frac{m}{2}\delta^{ab}\delta_{m+n,0} + i{\epsilon^{ab}}_c J^c_{m+n}
            \\
\com{J^a_m}{G^\alpha_{\dot{A},n}} &=& \frac{1}{2}{(\sigma^{aT})^\alpha}_\beta G^\beta_{\dot{A},m+n}
             \\
\ac{G^\alpha_{\dot{A},m}}{G^\beta_{\dot{B},n}} &=& \hspace*{-4pt}\epsilon_{\dot{A}\dot{B}}\bigg[
   (m^2 - \frac{1}{4})\epsilon^{\alpha\beta}\delta_{m+n,0}
  + (m-n){(\sigma^{aT})^\alpha}_\gamma\epsilon^{\gamma\beta}J^a_{m+n}
  + \epsilon^{\alpha\beta} L_{m+n}\bigg]\\
\com{L_m}{L_n} &=& \frac{m(m^2-\frac{1}{4})}{2}\delta_{m+n,0} + (m-n)L_{m+n}\\
\com{L_m}{J^a_n} &=& -n J^a_{m+n}\\
\com{L_m}{G^\alpha_{\dot{A},n}} &=& \left(\frac{m}{2}-n\right)G^\alpha_{\dot{A},m+n}.
\end{eqnarray}
\end{subequations}

\section{Useful Series}\label{ap:series}

We collect the series that we use throughout the paper here. These
series arise when considering a single boson and fermion.
\begin{align}\label{eq:id-1}
\sum_{k\,\text{odd}^+}\frac{1}{(n-\frac{k}{2})(k+l)}\frac{\Gamma(\frac{k}{2}+1)}
                                                         {\Gamma(\frac{k+1}{2})} 
 &= \frac{\pi l\Gamma(\frac{l+1}{2})}{4\Gamma(\frac{l}{2}+1)}\frac{1}{n+\frac{l}{2}}
      \left(\frac{\Gamma(\frac{l}{2})\Gamma(-n+\frac{1}{2})}{\Gamma(\frac{l+1}{2})\Gamma(-n)} 
              - 1\right)\\
\sum_{k\,\text{odd}^+} \frac{1}{k(p+k)(n-\frac{k}{2})}
         \frac{\Gamma(\frac{k}{2}+1)}{\Gamma(\frac{k+1}{2})}
 &= -\frac{\pi\Gamma(\frac{p+1}{2})}{4\Gamma(\frac{p}{2}+1)}\frac{1}{n+\frac{p}{2}}
    \left(
   \frac{\Gamma(\frac{p}{2}+1)\Gamma(-n+\frac{1}{2})}{\Gamma(\frac{p+1}{2})\Gamma(1-n)} -1\right)
\end{align}

These series arise when considering a single fermion:
\begin{subequations}
\begin{align}
\sum_{k\,\text{odd}^+}\frac{z_0^{n-\frac{k}{2}}\gamma^F_{pk}}{n-\frac{k}{2}}
 &=  \frac{z_0^{n+\frac{p}{2}}}{2(n+\frac{p}{2})}
  \left(\frac{\Gamma(\frac{p}{2}+1)\Gamma(-n+\frac{1}{2})}{\Gamma(\frac{p+1}{2})\Gamma(-n+1)} 
                   - 1\right)\\
\sum_{k\,\text{odd}^+}\frac{z_0^{n-\frac{k}{2}}\gamma^F_{kp}}{n-\frac{k}{2}}
 &=-\frac{z_0^{n+\frac{p}{2}}}{2(n+\frac{p}{2})}
  \left(\frac{\Gamma(\frac{p}{2})\Gamma(-n+\frac{1}{2})}{\Gamma(\frac{p+1}{2})\Gamma(-n)} 
                   - 1\right).\label{eq:gammaF-sum-1}
\end{align}
\end{subequations}

Some other useful series are
\begin{subequations}\begin{align}
\sum_{k\,\text{odd}^+}\frac{\Gamma(\frac{k}{2})}{\Gamma(\frac{k+1}{2})(m+\frac{k}{2})}
 &= \pi\frac{\Gamma(m+\frac{1}{2})}{\Gamma(m+1)}\\
\sum_{l\,\text{odd}^+}\frac{l}{(m-\frac{l}{2})(n-\frac{l}{2})}
      \frac{\Gamma(\frac{l}{2})}{\Gamma(\frac{l+1}{2})}
 &= 2\pi^2 m \frac{\Gamma(n)}{\Gamma(n+\frac{1}{2})}\delta_{m,n}\qquad m,n>0
\end{align}\end{subequations}

\bibliography{fuzzballs}

\end{document}